\documentclass[onecollarge,natbib]{svjour2}
\bibpunct{[}{]}{;}{n}{}{,} 
\smartqed  
\usepackage{graphicx}
\journalname{Few-Body Systems (EFB22)}
\begin{document}

\title{Few-body aspects of the near threshold pseudoscalar meson production
}


\author{Pawe{\l} Moskal         
}


\institute{P. Moskal,
              Institute of Physics, Jagiellonian University, Cracow, Poland, and 
              IKP, Forschungszentrum Juelich, Germany;
              Tel.: +48-12-6635558; 
              Fax: +48-12-6637086;
              \email{p.moskal@uj.edu,pl}           
}

\date{Received: date / Accepted: date}

\maketitle

\begin{abstract}
During last decade large samples of data have been collected on the production of the
ground-state pseudoscalar mesons in collisions of proton or deuteron beam with hydrogen or deuterium target.
These measurements have been performed in the vicinity of the kinematical threshold for meson production
where only a few partial waves in both initial and final state are expected to contribute
to the production process. This simplifies significantly the interpretation of the data,
yet still appears to be challenging due to the three or four particle final state systems
with a complex hadronic potential. We review experiments and phenomenology of the near
threshold production of the ground-state mesons in the few-body final states as for example:
nucleus-meson and  nucleon-nucleon-meson, and report on the status of the search of the mesic-nuclei
(a meson-nucleus bound states). 
Experimental advantages of  measurements close to the kinematical threshold are discussed,
and general features of the production mechanism of the $\eta$ and $\eta^{\prime}$ mesons in the nucleon-nucleon collisions
are presented emphasising results of  measurements of spin and izospin dependence of the production cross sections.

\keywords{meson production \and mesic nuclei \and  meson-nucleon interaction}
\end{abstract}

\section{Introduction}
\label{intro}
The organisers encouraged the author to summarize the threshold meson production studies conducted so far,
however due to the limitations in the number of pages this contribution cannot 
be treated as a summary. 
It is rather a very brief description of some of the studies on the meson production at threshold
reflecting the interest of the author. 
We will concentrate on the  $\eta$ and $\eta^{\prime}$ mesons and the report will be restricted to the hadronic processes
whereas for the photo-production the interested reader is referred e.g. to~\cite{ELSA-MAMI-plan-Krusche}.

Studies of the production of mesons in the collisions of hadrons are conducted in order to learn about the structure of these mesons,
their production mechanisms and about their hadronic interactions with nucleons~\cite{REVIEW,HAB,ACTA-Colin,wilkinjohansson}. 
Measurements and interpretation of results of hadronic processes in the energy regime of one GeV
is challenging since in the low energy Quantum Chromodynamics
the processes involving strong interaction cannot be treated perturbatively, 
and the analysis must rely on effective field theories 
and models. 
The $\eta$ and $\eta^{\prime}$ mesons are short lived and it is unfeasible to accomplish out of them a beam or target, and thus their interaction
with other hadrons cannot be investigated in the standard way via scattering experiments.
However, production of these mesons close to the kinematical threshold with low relative velocities to nucleons
gives a chance to study their interaction with nucleons 
which may manifest itself as structures in a meson-nucleon invariant mass distributions
and as enhancements in the cross section excitation functions with respect
to predictions based on the assumption
that the kinematically available phase space is  homogeneously populated. 

Interpretation of measurements near the threshold is  simplified due to the strongly suppressed contribution form the higher partial waves,
due to the low (tending to zero at threshold) 
relative momentum between the outgoing particles and because of small distances at which the creation of the meson accurs.
But nevertheless
it remains challenging to discern effects of the meson-nucleon interaction
from the overwhelming nucleon-nucleon interaction. 
This is why fifty years after the discovery of   
the $\eta$ and $\eta^{\prime}$ mesons~\cite{pevsner421,kalb64,gold64}
their hadronic interaction with nucleons is still not well known.
In the case of the $\eta^{\prime}$ meson its threshold production e.g. via the $pp\to pp\eta^{\prime}$ reaction
occurs at distances of the colliding nucleons in the order of 0.2~fm. Therefore, gluonic degrees of freedom may play a role 
in the production processes~\cite{GLUE-PL-Bass99}, implying that for the description of the meson production at threshold both 
hadronic
and quark-gluon degrees of freedom should be considered. 

Meson production at threshold is dominated by an S-wave in the final state but at the same time
threshold "filters" a single partial wave in the initial state. For example, 
protons collide predominantly in the $^3P_0$ state~\cite{REVIEW}
in the case of the pseudoscalar meson production in proton-proton
interactions.

Close-to-threshold the production cross section of pseudoscalar mesons
is reduced by the initial state interaction (ISI)
of the colliding nucleons by a factor of about 4~\cite{batinic321,hanhart176}
and it is enhanced by the final state interaction (FSI)
by more than an order of magnitude~\cite{REVIEW}.
The reduction of the cross section due to the ISI strongly depends on the meson mass, 
and therefore as it was introduced 
in reference~\cite{Moskal-Swave},
a strength of the production dynamics is at best expressed 
as the total cross section normalized to the ISI factor. A
natural variable for comparing the production dynamics for different mesons,
making it approximately independent of the meson mass, 
is the volume of the available phase space~\cite{Moskal-Swave}.
Such comparison was made for the $\pi^0$, $\eta$ and $\eta^{\prime}$ mesons
and it was shown that the dynamics for the $\eta$ meson production is about six time stronger 
than for the $\pi^0$ meson, which again is by further factor of six stronger than that of the $\eta^{\prime}$~\cite{Moskal-Swave}. 

A suppressed contribution of higher than $l = 0$ partial waves is of a great advantage for the measurements at threshold,
however a precise quantitative determination of their contributions is important especially for the studies of the meson nucleon interaction
which effects on the cross sections distributions are small and may be easly burdened by the effects caused by the higher partial waves.
E.g. enhancements in the meson-nucleon invariant mass distribution due to the meson-nucleon interaction
may be mixed with effects due to the p-wave nucleon-nucleon production~\cite{ETA-PRC-Moskal,ETAPRIME-Klaja}.
Experimental determination of contributions from various partial waves is difficult and requires
measurements with polarised beams and targets. 
So far such measurements were performed only for the $\pi^0$ meson production~\cite{meyer064002,meyer5439}.
In case of the $\eta$ mesons only beam analysing power $A_y$ for the $pp\to pp\eta$ reaction was studied 
with a poor precision~\cite{ETA-Ay-Czyzykiewicz,ETA-Ay-EPJ-Winter,ETA-Ay-PL-Winter,ETA-Ay-Balestra}
where the best result of $A_y$ determined for four angular bins with uncertainty of~$\pm 0.1$
is based on about 2000 reconstructed events~\cite{ETA-Ay-Czyzykiewicz}.
In the near future, the precision of the determination of angular dependence of the analysing power for the $pp \to pp\eta$ reaction
will be improved up to $\pm 0.01$
based on a new high statistics data sample (about $10^6$ reconstructed $pp \to pp\eta$ events)
collected using the large acceptance and azimuthally symmetric WASA detector and a polarized proton beam of
the Cooler Synchrotron COSY~\cite{ETA-Ay-WASA-Moskal,ETA-Ay-WASA-Hodana}. 
However, the author is not aware of any plans of measurements of the spin observables 
for the $\eta^{\prime}$ meson production 
in the collissions of nucleons.
%
%
\section{Experimental advantages of the threshold meson production}
In the case of the fixed target experiments the fast movement of the center-of-mass system along the beam line
cause, for the close to threshold reactions, that all ejectiles are confined 
in a narrow cone and can be efficiently registered with the detectors of relatively 
small sizes. The emission of all ejectiles under small angles 
enables to use dipole magnets as  
charged particles analyzers for momentum determination with high precision
in the whole available phase space volume.
A description of the typical zero-degree facilities dedicated 
to the studies of the threshold meson production can be found e.g. 
in~\cite{COSY-11-Brauksiepe,COSY-11-Klaja,ANKE}.
In addition, in the case of the missing mass techique used 
for the identification of the produced mesons,
a 
measurement at threshold improves mass resolutions because 
at threshold the partial derivative of the missing mass
with respect to the momentum of outgoing ejectile tends to zero~\cite{Czerwinski-PRL}. 
Moreover,
close to threshold
the signal-to-background ratio increases due to the more rapid reduction of the phase space  for
multimeson production than for~the~single meson~\cite{Czerwinski-PRL}.

As examples of successful measurements benefiting from the threshold kinematics it is worth to mention:
 (i) Determination 
             of the natural width of $\eta^{\prime}$ meson directly from its mass distribution for the first time with the resolution comparable
    to its width ($\sim$200~KeV)~\cite{Czerwinski-PRL}. 
    The precision of this single experiment~\cite{Czerwinski-PRL} was equal to the accuracy which Particle Data Group had previously achieved
    by combining  51 experiments which determined properties of the eta-prime meson only indirectly connected with its width~\cite{PDG2008},
   (ii) Determination of the 
    value of the mass of the $\eta$ meson~\cite{ANKE-ETAmass} with high precision based on the close-to-threshold measurement of the $dp\to ^3\!\!He \eta$ reation 
    where the $\eta$ meson was identified via 
    missing mass distribution and 
    the individual beam momenta were fixed with a relative precision of $dp/p \sim 3 \cdot 10^{-5}$ using 
a polarized deuteron beam of COSY and inducing an artificial depolarizing spin resonance~\cite{COSY-depolarizing}.

The above discussed advantages of the meson production at threshold are, however,  valid only for studies of mesons with 
a narrow spectral functions. In the the case of broad resonances 
(as e.g. $f_0$ or $a_0$)
the notion close-to-threshold becomes non-trivial.
In this case a phase space volume varies significantly within the mass range of the meson 
and the determination of cross section
requires a scanning of the excess energies in the range of the spectral function of the studied meson.
A formal definition of the close-to-threshold
total cross section for broad resonances was first introduced in reference~\cite{Moskal-f0a0JPG},
where it was applied to the interpretation
of the threshold production of the $f_0$ meson in the collisions of protons~\cite{Moskal-f0a0JPG}.


\section{Threshold production of the $\eta$-nucelon-nucleon and $\eta^{\prime}$-nucleon-nucleon systems}
The near threshold production of the $\eta$ meson was intensively studied
in nucleon-nucleon collisions for which a total~\cite{ETABergdolt,ETAChiavassa,ETACalen,ETA-ETAPRIME-Hibou,ETASmyrski}
and differential~\cite{ETA-PRC-Moskal,ETA-Abdel,ETA-EPJ-Moskal,ETA-Petren} cross sections have been established for the $pp\to pp\eta$ reaction
as well as for the $pn \to pn \eta$ process~\cite{pn-deta-Calen,pn-pneta-Calen,pn-pneta-Moskal}. In case of the $pp\to pp\eta$ reaction
an angular dependence of the analysing power $A_y$ has also been 
established~\cite{ETA-Ay-Czyzykiewicz,ETA-Ay-EPJ-Winter,ETA-Ay-PL-Winter,ETA-Ay-Balestra}.
It is important to stress that results from different laboratories (CELSIUS, COSY, SATURNE) 
are consistent within 
the estimated systematical uncertainties which are in the order of 10\%.

The poroton-neutron reaction were realized via quasi-free proton-neutron collisions using a dueteron as a source of neutrons.
In the data analysis it was assumed that 
a spectator proton leaves the deutron 
undisturbed and that 
it is on mass shell already at the collision moment
and that the matrix element
for the production of the $\eta$ meson by the  beam proton off the neutron  bound in the deuteron
is identical to that for the free $pn\to pn\eta$ reaction~\cite{REVIEW,HAB}.
These assumptions are well supported by a theoretical considerations~\cite{kaptari}
and were confirmed by many experiments~\cite{pn-deta-Calen,pn-pneta-Moskal,quasi-free-tlo-method-Moskal,tofpn1,tofpn2,triumf}
which 
have proven that spectator model is valid at least for Fermi momentum
up to 150 MeV/c~\cite{tofpn1}.

A strong dependence of the $\eta$ meson production on the izospin of the colliding nucleons was observed.
Total cross sections for the quasi-free $pn\to pn\eta$
exceed
corresponding cross sections for the $pp\to pp\eta$ reaction  by a factor of about
three at thereshold~\cite{pn-pneta-Moskal} and by factor of six 
at higher excess energies~\cite{pn-pneta-Calen}.
Combining information of the strong isospin dependence and the isotropic angular distributions of the $\eta$ meson emission angle in the center-of-mass frame,
it was established that the $\eta$ meson is predominantly created via excitation of one of the nucleons 
to the $N^*(1535)$ resonance via exchange of the isovector meson~\cite{wilkinjohansson}
and the angular dependence of the analysing power slightly indicated that the process proceed 
via exchange of the $\pi$ meson~\cite{ETA-Ay-Czyzykiewicz}.


In the case of the $\eta^{\prime}$ meson production in the collisions of nucleons
almost 40 times smaller cross sections were observed~\cite{ETA-ETAPRIME-Hibou,ETAPRIME-Moskal,ETAPRIME-PL-Moskal,ETAPRIME-Balestra,ETAPRIME-Khoukaz,ACTA-Eryk} 
than in the case of the $\eta$ meson, which indicates that in contrast to the $\eta$ meson the $\eta^{\prime}$ is produced non-resonantly.
Based on the comparison of the shape of the excitation functions for the $pp\to pp\eta$ and $pp\to pp\eta^{\prime}$ reactions it was concluded that 
the $p-\eta$ interaction is much stronger than the $p-\eta^{\prime}$~\cite{Moskal-Swave}.
At first, the small values of the cross sections for the $pp\to pp\eta^{\prime}$ reaction~\cite{ETAPRIME-Moskal} 
were even interpreted as an indication of the 
$p-\eta^{\prime}$ repulsive interaction~\cite{ETAPRIME-Baru}. This hypothesis was however excluded later by the more precise data~\cite{ETAPRIME-PL-Moskal}.

Up to now, as regards the interaction of pseudoscalar mesons with nucleons the only well known is this of $\pi^0$ for which
a real part of the scattering legth is equal to 0.1294$\pm$0.0009~fm~\cite{pi0-Sigg}. The values of the real part of the $p-\eta$ 
scattering length are known much less precisely and varies from 0.20~fm to 1.05~fm
depending on the analysis method~\cite{greenR2167,N.Kaiser-II,green053,green035208}.
Differences in the value of $\eta N$ scattering lengths obtained in different analyses are at least to some extent
explained by the recent observation that the flavour singlet component induces greater binding 
than the flavour-octet one~\cite{GLUE-ACTA-Bass10,Mesic-Bass}.
Therefore, the $\eta-\eta^{\prime}$ mixing, which is neglected in many of the former analyses,
increase the $\eta$-nucleon scattering length relative
to the pure octet $\eta$ by a factor of about 2~\cite{Mesic-Bass}.
For the sake of completness, it is important to stress that 
based on the close to threshold cross sections measured for the $pp\to ppK^+K^-$ 
reaction~\cite{COSY11-Wolke,disto,COSY11-Quentmeier,COSY11-Winter,ANKE-Maeda,COSY11-Silarski,ANKE-Ye2012,ANKE-Ye2013,dzyubaKK} 
the scattering length and effective range of the interaction between strange pseudoscalar mesons $K^+K^-$ 
were recently estimated~\cite{COSY11-Silarski,SilarskiKK}. The resultant values are consistent with zero 
with rather large uncertaintes.

One of the interesting, and still not fully understood observations from the threshold production of $\eta$ and $\eta^{\prime}$ mesons, are large enhancements 
in invariant masses of two-particle subsystems seen both in the $pp\eta$~\cite{ETA-PRC-Moskal,ETA-Abdel,ETA-EPJ-Moskal}
and $pp\eta^{\prime}$~\cite{ETAPRIME-Klaja} systems.
Since the enhancement is similar in $\eta$ and $\eta^{\prime}$ case~\cite{ETAPRIME-Klaja},
and the strength of proton-$\eta$ and proton-$\eta^{\prime}$ interaction
seems to be different \cite{Moskal-Swave}, one can conclude that the observed enhancement
is not caused by a proton-meson interaction, especially that
calculations assuming a significant contribution of P-wave in the
final state~\cite{kanzo},
and models including
energy dependence of the production amplitude~\cite{deloff,ceci}, reproduce the data
within error bars.
Therefore, the determination of the spin observables and extraction of the contribution
from the higher partial waves is mandatory for the understanding of the observed enhancements 
in the discussed invariant mass spectra.

It is also instructive to determine the isospin dependece of the production cross sections which enables to disentangle contributions
from various meson exchanges in the reaction mechanisms and at the same time enables to learn about the structure of the produced mesons.
In this context especially interesting is the gluonium content of the $\eta^{\prime}$ meson which 
is discussed comprehensively in articls~\cite{GLUE-PL-Bass99,GLUE-ACTA-Bass10,GLUE-PL-Bass06,GLUE-Bass-Scripta},
where it is argued 
that
a comparison of the close-to-threshold total cross sections for the $\eta^{\prime}$
meson production in both the $pp\to pp\eta^{\prime}$ and $pn \to pn \eta^{\prime}$ reactions
should provide  insight into the flavour-singlet (perhaps also into gluonium)
content of the $\eta^{\prime}$ meson and the relevance of quark-gluon
or hadronic degrees of freedom in the creation process.
The production through the
colour-singlet object as suggested in reference~\cite{GLUE-PL-Bass99} is isospin independent
and should lead to the same production yield of the
$\eta^{\prime}$ meson in the $pn\to pn\ gluons \to pn\eta^{\prime}$
and                          $pp\to pp\ gluons \to pp\eta^{\prime}$ reactions
after correcting for the final and initial state interaction between the nucleons~\cite{HAB}.
In the case of the $\eta$ meson production
in collisions of nucleons
the creation from isospin I~=~0
exceeds the production with I~=~1
by a factor of about 12, indicating production through the exchange of isovector objects. 
However in case of the $\eta^{\prime}$ meson 
so far only an upper limits of cross sections for I~=~0 have been established~\cite{ETAPRIMEpn-Klaja}.
The result indicates weaker isospine dependence for the $\eta^{\prime}$ meson production with respect to the $\eta$ meson,
and disfavours the dominance of the $N^*(1535)$
resonance in the production process of the $\eta^{\prime}$ meson.
However the so far achieved accuracy is not suffcient to conclude about 
a gluonium content in the $\eta^{\prime}$ meson.

\section{Threshold production of the $\eta-^3\!\!He$ and $\eta-^4\!\!He$ systems and the search for the eta-mesic helium}
The production of the $\eta$ meson has been also studied intensively in the proton-deuteron and deutron-deutron reactions e.g.:
$pd \to ^3\!\!He \eta$~\cite{3Heeta-Berger,3Heeta-Mayer,3Heeta-Betigeri,3Heeta-Adam,3Heeta-Acta-Smyrski,3Heeta-PL-Smyrski,3Heeta-Mersmann},
$dd \to ^4\!\!He \eta$~\cite{4Heeta-Frascaria,4Heeta-Willis,4Heeta-Wronska,4Heeta-Budzanowski}, and
$pd \to  pd \eta$~\cite{pdeta-Hibou,pdeta-Bilger,pdeta-Cezary}. Similarly as in the case of the nucleon-nucleon-meson final state
a large enhancement of the cross sections over the predictions based on the assumption of the homogeneously 
populated phase space was observed at threshold
for all listed reactions. 
The steepest rise of the total cross section is seen for the $pd \to ^3\!\!He \eta$ reaction. 
It grows from zero up the the value of about 400~nb over the range of 1 MeV of excess energy and next keeps almost 
in the excess energy range of  about 10~MeV.
This enhancement may be assigned to the $^3\!He-\eta$ interaction,
because it is also observed in the photo-production reactions~\cite{KruscheActa,gamma3He-Pfeiffer,gamma3He-Pheron},
thus it is independent on the initial channel,
and because the s-wave production amplitude is fairly energy independent~\cite{3Heeta-Mayer,WilkinActa,3Heeta-t20-Khoukaz}.
Moreover, 
the asymmetry in the angular distribution of the $\eta$ meson emission~\cite{3Heeta-PL-Smyrski,3Heeta-Mersmann}
    indicates strong changes of the phase of the s-wave production amplitude with energy,
    as expected from the occurrence of the
    bound or virtual $\eta {^{3}\mbox{He}}$ state~\cite{wilkin2}. Again similar behaviour is observed in photo-production reactions
    where evolution  of the angular dependence of~$\gamma ^3\!He \to \eta ^3\!He$~\cite{gamma3He-Pheron} as a function of energy,
    indicates changing of s-wave amplitude
    associated with the $\eta{^{3}\mbox{He}}$ pole~\cite{WilkinActa}.

Existence of a bound state of the $\eta$ meson and nucleus (referred to as a mesic nucleus) 
was predicted 28 years ago~\cite{Haider}.
Initially
the $\eta$--mesic nuclei were considered
to exists for A~$\ge$~12~only~\cite{Haider}
due to the
relatively small value of the $\eta$N scattering length estimated in eighties~\cite{liu1}. 
A decade later, a large values of the $\eta$-nucleon scattering length (1~fm)
were extracted in some analysis reported in~\cite{wycech} 
which do not exclude the formation
of  bound $\eta$-nucleus states  for such light nuclei as helium~\cite{wycech1,Colin-mesic}
or even for deuteron~\cite{green}.
However, so far none of the experiments have confirmed univocally the existence of such a state
neither in 
reactions induced by pions~\cite{Mesic-bnl}, 
protons~\cite{Mesic-gem}, deuterons~\cite{COSY11-MoskalSmyrski,Mesic-Adlarson,Mesic-Afanasiev-JINR} or photons~\cite{gamma3He-Pheron,Mesic-Baskov-LPI}.
In the searches for the direct signal from the $\eta$-mesic helium  the established so far upper limits amount to
about 270~nb
for the 
 $dp \to ({^{3}\mbox{He}} \eta)_{bound} \to ppp\pi^-$  reaction~\cite{COSY11-MoskalSmyrski},
about 70~nb
for the
 $dp \to ({^{3}\mbox{He}} \eta)_{bound} \to {^{3}\mbox{He}} \pi^0$  reaction~\cite{COSY11-MoskalSmyrski},
and
about 25~nb
for the 
 $dd \to ({^{4}\mbox{He}} \eta)_{bound} \to {^{3}\mbox{He}} p\pi^-$  reaction~\cite{Mesic-Adlarson}. 
The determined upper limits are close to the newly predicted values of total cross sections
for the dd and pd reaction at the $\eta$-mesic pole~\cite{WilkinActa,Wycech-Acta}, which amounts to 
about 80~nb for the $pd \to ({^{3}\mbox{He}}-\eta)_{bound} \to X p \pi^-$~\cite{WilkinActa}
and is in the range from 4.5~nb~\cite{Wycech-Acta}
to 30~nb for the $dd \to (4He-eta)_{bound} \to X p \pi^-$ reaction~\cite{WilkinActa}.
Although so far not successful, the experimental~\cite{ELSA-MAMI-plan-Krusche,WASA-at-COSY-SkuMosKrze,eta-mesic-JPARC-Fujioka,Mesic-Nucletron-Afanasiev} 
and 
theoretical~\cite{ACTA-Colin,GLUE-ACTA-Bass10,Mesic-Bass,Wycech-Acta,Wycech-Acta10,ETA-Friedman-Gal,ETA-Gal-Cieply,Mesic-Kelkar,
EtaMesic-Hirenzaki} 
investigations of the $\eta$-mesic nucleus 
are being continued. The observation of such a bound state and determination of its properties would be very valuable for the 
determination of the $\eta$-nucleon interaction,
the N$^*(1535)$ properties in nuclear matter~\cite{EtaMesic-Hirenzaki,jido},
the properties of the $\eta$ meson
in the nuclear medium~\cite{wycech,ETA-Friedman-Gal,osetNP710},
and in general the studies of the chiral and axial U(1) symmetry breaking in low energy QCD~\cite{Mesic-Bass,GLUE-PL-Bass06,EtaMesic-Hirenzaki}.
The properties of $\eta$ and $\eta^{\prime}$ mesic nucleus are strongly 
sensitive to the contribution of the flavour-singlet component of these mesons.  
Therefore, the $\eta^{\prime}$-mesic nucleus is very interesting in this context too.
  
The quark condesate is modified in nucleus which changes the properties of hadrons in nuclear medium.
  The binding energies of $\eta$ and $\eta^{\prime}$ in medium are sensitive to the non-perturbative glue
  associated with the axial U(1) dynamics~\cite{GLUE-PL-Bass06,Mesic-Bass}, and as stated in~\cite{Hirenzaki-ACTA} "due to 
  the UA(1) anomaly effect, a relatively large mass reduction of $\eta^{\prime}$ meson is expected
  at nuclear saturation density, which may indicate the existence of the $\eta^{\prime}$-mesic nucleus".
However, so far most of the experimental studies have been concentrated on the search for the $\eta$-mesic nuclei  because the $\eta$-nucleon 
interaction seems to be much stronger than the $\eta^{\prime}$-nucleon or $\pi$-nucleon~\cite{Moskal-Swave}.
Yet, recently there are vigourous 
theoretical~\cite{Mesic-Bass,EtaMesic-Hirenzaki,eta-prime-mesic-Nagahiro,eta-prime-mesic-Nagahiro-Oset,eta-prime-mesic-Nagahiro-Jido}
and experimental~\cite{ELAS-BGO-OD-Metag,eta-prime-mesic-GSI-tanaka,eta-prime-mesic-GSI-Itahashi}
investigations
of feasibility of the observation of the $\eta^{\prime}$-mesic 
nuclei started by the predictions published in~\cite{eta-prime-mesic-Hirenzaki}. 
Based on the cross sections from the $pp\to pp\eta^{\prime}$ reactions a scattering length of the $p-\eta^{\prime}$ 
potential seems to be small~\cite{Moskal-Swave}, on the other hand recent
photoproduction measurements of CBELSA/TAPS~\cite{Nanova2012,Nanova2013} 
shows that the 
real part of the $\eta^{\prime}$-nucleus optical potential is larger than  its imaginary part giving 
a hope for the observation of the $\eta^{\prime}$ mesic nucleus. 

The search for the $\eta$ and $\eta^{\prime}$ mesic nucleus is exciting and there are plans to continue this investigations in the future.

\begin{acknowledgements}
We acknowledge support
by the Polish National Science Center through grants No. 0320/B/H03/2011/40, 2011/01/B/ST2/00431, 2011/03/B/ST2/01847,
by the FFE grants of the Research Center J\"{u}lich, by the Foundation for Polish Science (MPD programme),
by the EU Integrated Infrastructure Initiative HadronPhysics Project under contract number RII3-CT-2004-506078,
and by the European Commission under the 7th Framework Programme through the 
’Research Infrastructures’ action of the ’Capacities’ Programme, Call: FP7~-~INFRASTRUCTURES~-~2008~-~1, Grant Agreement N. 227431,
\end{acknowledgements}



\end{document}